\documentstyle[12pt]{article}
\newcommand{\Ir}{Z\!\!\!Z}
\newcommand{\idty}{{\leavevmode{\rm 1\mkern -5.4mu I}}}
\newcommand{\Ibb}[1]{ {\rm I\ifmmode\mkern
            -3.6mu\else\kern -.2em\fi#1}}
\newcommand{\ibb}[1]{\leavevmode\hbox{\kern.3em\vrule
     height 1.2ex depth -.3ex width .2pt\kern-.3em\rm#1}}
\newcommand{\Cx}{{\ibb C}}
\newcommand{\Nl}{{\Ibb N}}
\newcommand{\Rl}{{\Ibb R}}
\begin{document}
\renewcommand{\theequation} {\arabic{section}.\arabic{equation}}

\begin{tabbing}
\hspace*{12cm}\= GOET-TP 96/95 \\
              \> September 1995
\end{tabbing}
\vskip1.cm

\centerline{\huge \bf Umbral Calculus, Discretization,}
\vskip.5cm
\centerline{\huge \bf and Quantum Mechanics on a Lattice}

\vskip1.cm
\begin{center}
\begin{minipage}{13cm}
{\bf A. Dimakis}$^\dagger$, {\bf F. M\"uller-Hoissen}$^\ddagger$
             \ and \ {\bf T. Striker}$^\ddagger$
\vskip .3cm
$^\dagger$ Department of Mathematics, University of Crete,
GR-71409 Iraklion   \\
$^\ddagger$ Institut f\"ur Theoretische Physik,
Bunsenstr. 9, D-37073 G\"ottingen
\end{minipage}
\end{center}
\vskip1.cm

\begin{abstract}
\noindent
`Umbral calculus' deals with representations
of the canonical commutation relations. We present a short exposition
of it and discuss how this calculus can be used to discretize continuum
models and to construct representations of Lie algebras on a lattice.
Related ideas appeared in recent publications and we show that the
examples treated there are special cases of umbral calculus. This
observation then suggests various generalizations of these examples.
A special umbral representation of the canonical
commutation relations given in terms of the position and momentum
operator on a lattice is investigated in detail.
\end{abstract}

\section{Introduction}
\setcounter{equation}{0}
Umbral calculus\footnote{This terminology goes back to
the nineteenth century mathematician Sylvester who used the Latin
word {\em umbra} to denote something which would nowadays be called a
linear functional. See also \cite{Roma+Rota78}.}
is an analysis of certain representations of the commutation relations
\begin{eqnarray}             \label{ccr}
  \lbrack Q_i \, , \, \hat{\bf x}_j \rbrack = \delta_{ij} \, \idty
  \; , \qquad
  \lbrack Q_i \, , \, Q_j \rbrack = 0 = \lbrack \hat{\bf x}_i \, , \,
                                        \hat{\bf x}_j \rbrack
\end{eqnarray}
in terms of operators on the algebra of polynomials in variables
$x_i, \, i=1, \ldots,n$ (see \cite{Roma+Rota78,RKO73} for reviews).
In particular, it provides us with representations by operators
acting on polynomials of {\em discrete} variables. Let us assume
that $Q_i, \hat{\bf x}_j$ is such a representation and let
$A(y_i, \partial/\partial y_j) \, f(y_k) = 0$
be a differential equation on $\Rl^n$ with a polynomial solution
$f$.\footnote{Here and in the following an expression like $f(y_k)$
stands for $f(y_1,\ldots,y_n)$.}
Introducing multiplication operators ${\bf y}_i$, we can write
it in the form
\footnote{In this expression the $1$ plays the role of a `state' on
which we act with an operator algebra to generate an irreducible
representation of the latter.}
\begin{eqnarray}
   A({\bf y}_i, \partial/\partial y_j) \, f({\bf y}_k) \, 1 = 0 \; .
\end{eqnarray}
The operators ${\bf y}_i$ and $\partial/\partial y_j$ do satisfy the
commutation relations (\ref{ccr}), of course.
The verification that $f(y_k)$ solves the original differential
equation is now translated into an algebraic problem which only
requires the abstract commutation relations (\ref{ccr}), i.e., it
does not depend on the specific choice of representation. Defining
$ \tilde{f}(x_k) := f(\hat{\bf x}_k) \, 1 $, then also
\begin{eqnarray}
        A(\hat{\bf x}_i, Q_j) \, \tilde{f}(x_k) = 0
\end{eqnarray}
holds which is a {\em difference} equation. We have simply substituted
\begin{eqnarray}          \label{umap}
   {\bf y}_i \quad \mapsto \quad \hat{\bf x}_i \; ,  \qquad \quad
   {\partial \over \partial y_j}  \quad \mapsto  \quad Q_j   \; .
\end{eqnarray}
If $f(y_k)$ solves the original differential equation, then $\tilde{f}
(x_k)$ is a solution of the corresponding difference equation.
\vskip.2cm

For differential equations possessing polynomial solutions,
the notion of quasi-exact solvability has been introduced \cite{Turb92}.
Several examples are provided by eigenvalue problems in quantum
mechanics. A corresponding example for the above discretization
procedure appeared recently in \cite{ST95}. In section
\ref{isospectral} we show that its underlying structure is umbral
calculus.
\vskip.2cm

The above operator substitution yields a mapping of an eigenvalue
equation for a differential operator to an eigenvalue equation for a
difference operator together with a `formal' mapping of solutions. It
seems that we have a general procedure for `isospectral
discretization' of differential operator eigenvalue problems.
The problem, however, is that (besides for polynomials) the mapping of
solutions in general only works at the level of formal power series,
but does not respect convergence properties.
\vskip.2cm

Also, in the abovementioned treatment \cite{ST95} of
eigenvalue problems one does not really get a discretization of the
original quantum mechanical problem since that involves
non-polynomial functions. For serious applications we therefore
need an extension of the procedure sketched above beyond polynomials
and formal power series. Such a discretization method could then be
of interest for solving differential equations numerically.
\vskip.2cm

The commutation relations of differential operators $A({\bf y}_i,
\partial/\partial y_j)$ and $B({\bf y}_i, \partial/\partial y_j)$
are preserved under the substitution (\ref{umap}).
In this way one obtains representations of operator algebras, in
particular Lie and Hopf algebras, by operators acting on functions on
a lattice. An example appeared recently in \cite{FS95} where
representations of the Poincar\'e and the $\kappa$-deformed Poincar\'e
algebra \cite{LNR92} on a lattice were constructed. In section
\ref{modUC} we explain how it fits into the umbral framework.
\vskip.2cm

All this raises the question whether it is possible to understand
(some of the) umbral maps (\ref{umap}) on algebras of
non-polynomial functions. In view of possible applications
to quantum mechanics, it would be of interest to have $Q_i,
\hat{\bf x}_j$ defined on the Hilbert space of square summable
functions on a lattice.
Is it possible that $\hat{\bf x}_i$ and $-{\rm i} Q_j$ (which as a
consequence of (\ref{ccr}) satisfy the canonical commutation relations
of quantum mechanics) are selfadjoint operators and is (\ref{umap})
perhaps a unitary equivalence ?
Our work intends to contribute to the clarification of such questions.
An example of particular interest is suggested by the work in
\cite{FS95}. The representation of the canonical commutation relations
which appeared there is investigated in detail in section \ref{QM}.
\vskip.2cm

Section \ref{IntroUC} contains a brief introduction to our understanding
of umbral calculus. By no means it intends to cover the whole subject.
An example treating symmetries on a lattice is then presented in section
\ref{Solatt}. Another application is discussed in section
\ref{isospectral}, partly motivated by \cite{ST95}.
In section \ref{modUC} we slightly generalize the umbral framework of
section \ref{IntroUC}. We also comment on a representation of the
Poincar{\'e} algebra on a lattice which appeared in \cite{FS95}.
Its underlying representation of the canonical commutation relations
is the subject of section \ref{QM}. It leads us to a framework for
quantum mechanics on a lattice.
Some conclusions are collected in section \ref{Conclusions}.

\section{A brief introduction to umbral calculus}
\label{IntroUC}
\setcounter{equation}{0}
In this section we recall some notions and results from umbral
calculus. We refer to \cite{RKO73,Roma+Rota78} for the
corresponding proofs and further results.
For simplicity, we restrict our considerations to the case of a single
`coordinate' $x$. All results extend to several (commuting) variables in
an obvious way.
\vskip.2cm

An operator $O$ acting on the algebra (over a field of characteristic
zero, like $\Rl$ or $\Cx$) of polynomials in $x$ is
{\em shift-invariant} if it commutes (for all $a$ in the field) with
the shift operators $S_a$ (defined by $S_a f(x) = f(x+a)$).
\vskip.2cm

The {\em Pincherle derivative} of an operator $O$ is defined as
the commutator
\begin{eqnarray}
  O' := \lbrack O \, , \, {\bf x} \rbrack = O \, {\bf x} - {\bf x} \, O
\end{eqnarray}
where $\bf x$ is the multiplication operator, acting on polynomials
in $x$ by multiplication with $x$.
The Pincherle derivative of a shift-invariant operator is again a
shift-invariant operator. The {\em umbral algebra} is the algebra of
all shift-invariant operators. The Pincherle derivative is a derivation
of the umbral algebra.
\vskip.2cm

A {\em delta operator} $Q$ is a linear operator, acting on the algebra
of polynomials in $x$, which is shift-invariant and for which $Qx$ is a
nonzero constant. It can be shown that ${Q'}^{-1}$ exists (as a
linear operator on the space of polynomials) and commutes with $Q$. If
we define
\begin{eqnarray}        \label{umbral-hatx}
               \hat{\bf x} := {\bf x} \, {Q'}^{-1}
\end{eqnarray}
it follows that
\begin{eqnarray}               \label{ccr1}
         \lbrack Q \, , \, \hat{\bf x} \rbrack = \idty
\end{eqnarray}
where $\idty$ stands for the identity operator. In this way each delta
operator $Q$ provides us with a representation of the canonical
commutation relations on the algebra of polynomials in $x$.
\vskip.2cm

A {\em polynomial sequence} $q_k(x)$, $k=0,1,2 \ldots$, is a sequence
of polynomials where $q_k(x)$ is of degree $k$. A polynomial
sequence is called {\em basic} for a delta operator $Q$ if
$q_0(x) = 1$, $q_k(0) = 0$ whenever $k>0$, and
\begin{eqnarray}              \label{basic_sequence}
        Q \, q_k = k \, q_{k-1}    \; .
\end{eqnarray}
It turns out that basic sequences are of {\em binomial type},
i.e., they satisfy
\begin{eqnarray}
 q_k(x+y) = \sum_{\ell=0}^k \, {k \choose \ell} \, q_\ell(x) \,
            q_{k-\ell}(y) \;.
\end{eqnarray}
The basic polynomial sequence for $Q$ is given by
\begin{eqnarray}      \label{Rodrigues}
        q_k(x) = \hat{\bf x} \, q_{k-1}(x) = \hat{\bf x}^k \, 1
\end{eqnarray}
which is known as the {\em Rodrigues formula}.
\vskip.2cm

An operator which maps a basic polynomial sequence into another basic
polynomial sequence is called an {\em umbral operator} (\cite{RKO73},
p.28). Defining
\begin{eqnarray}      \label{f-tilde}
         \tilde{f}(x) := f(\hat{\bf x}) \, 1
\end{eqnarray}
for a polynomial $f$, (\ref{Rodrigues}) shows that the operator
$\,\tilde{} \,$ is an umbral operator.
\vskip.2cm

An associative and commutative product is defined by
\begin{eqnarray}
  \tilde{f}(x) \ast \tilde{h}(x) := f(\hat{\bf x}) \, h(\hat{\bf x})
  \, 1  \; .
\end{eqnarray}
In particular, $q_k(x) \ast q_\ell (x) = q_{k+\ell}(x)$. The delta
operator $Q$ is a derivation with respect to the $\ast$-product, i.e.,
\begin{eqnarray}
  Q \lbrack p(x) \ast q(x) \rbrack
                   = (Q p(x)) \ast q(x) + p(x) \ast Q q(x)
\end{eqnarray}
for polynomials $p$ and $q$.
\vskip.2cm
\noindent
{\em Example 1.}
For $Q = d/dx$ we have $Q'= \idty$ and therefore $q_k(x) = x^k$ which
is the simplest polynomial sequence.  \hfill {\Large $\Box$}
\vskip.2cm
\noindent
{\em Example 2.}
Let $Q=D/(D-1)$ with $D := d/dx$. Then $Q' = - (D-1)^{-2}$ and $q_k(x) =
\lbrack - {\bf x} \, (D-1)^2 \rbrack^k \, 1$ are the {\em basic Laguerre
polynomials} \cite{RKO73}.             \hfill {\Large $\Box$}
\vskip.2cm
\noindent
As pointed out in the introduction, we are particularly interested in
the case where the algebra of polynomials in $x$ can be realized as
an algebra of functions on a discrete set. In the following two examples
we may choose $x$ to be the canonical coordinate function on an infinite
lattice with spacings $a$ (where $a$ is a positive real number). In the
way outlined in the introduction, both examples provide us with a
prescription to translate functions on $\Rl$ and differential operators
into corresponding functions and operators on a lattice. The interesting
aspect is that this prescription not only maps a differential equation
into a corresponding difference equation, but it also allows us, in
principle, to calculate the solutions of the difference equation from
those of the differential equation (see sections \ref{isospectral} and
\ref{QM}).

\vskip.2cm
\noindent
{\em Example 3.}
Let $Q = \partial_+$ where $\partial_+$ is the {\em forward discrete
derivative} operator,
\begin{eqnarray}
   (\partial_+ f)(x) = {1 \over a} \lbrack f(x+a) - f(x) \rbrack
\end{eqnarray}
acting on a function $f$. We find $(Q'f)(x) = f(x+a)$ and therefore
$Q'=S_a$, the shift operator. Hence,
\begin{eqnarray}
            q_k(x) = ({\bf x} S_a^{-1})^k \, 1
                   = x (x-a) \cdots (x - (k-1) \, a )
                   = x^{(k)}
\end{eqnarray}
where $x^{(k)}$ is the $k$-th (falling) {\em factorial
function}.\footnote{For $a=1$ and $x \in \Nl$ it counts the number
of injective maps from a set of $n$ elements to a set of $x$ elements.}
Some formulas for the $\ast$-product associated with the discrete
derivative delta operator can be found in the appendix. Analogous
formulas hold for the {\em backward} discrete derivative operator
$\partial_-$ which is formally obtained from $\partial_+$ replacing $a$
by $-a$.               \hfill {\Large $\Box$}

\vskip.2cm
\noindent
{\em Example 4.}
For the {\em central difference operator} \cite{Rior}
\begin{eqnarray}       \label{centraldiff}
      Q = { 1 \over 2 a } \, ( S_a - S_{-a} )
        =  {1 \over 2} \, (\partial_+ + \partial_-)
\end{eqnarray}
we have $Qf(x) = \lbrack f(x+a) - f(x-a) \rbrack/(2a)$. Solving
(\ref{basic_sequence}), one finds the basic sequence
\begin{eqnarray}
    q_k(x) = x \, \prod_{n=1}^{k-1} (x+ka-2na)   \qquad (k > 1)
\end{eqnarray}
and $q_0(x)=1, \, q_1(x)=x$. Furthermore,
\begin{eqnarray}
   Q'  = { 1 \over 2 } \, ( S_a + S_{-a} )   \, , \qquad
   Q'' = a^2 \, Q   \; .            \label{central-Q'}
\end{eqnarray}
Using the Rodrigues formula,
\begin{eqnarray}
  Q'^{-1} q_k(x) = \prod_{n=1}^{k} \lbrack x + (k+1) a - 2 n a \rbrack
  \qquad (k > 0)
\end{eqnarray}
which shows that $Q'^{-1}$ is indeed well-defined on polynomials in $x$.
The operator $Q'^{-1}$ also exists as a selfadjoint operator in the
Hilbert space $\ell_2(a \Ir)$, see section \ref{QM}.
                            \hfill {\Large $\Box$}
\vskip.2cm
\noindent
{\em Example 5.} Over a finite field there are finite-dimensional
representations of the commutation relation (\ref{ccr1}).
Over $\Ir_3$ the matrices
\begin{eqnarray}
   {\bf x} = \left( \begin{array}{ccc} 0 & 0 & 0 \\
                                       0 & 1 & 0 \\
                                       0 & 0 & 2 \end{array} \right )
       \; , \qquad
   Q = \left( \begin{array}{ccc} 0 & 1 & 0 \\
                                 0 & 0 & 1 \\
                                 1 & 0 & 0 \end{array} \right )
\end{eqnarray}
provide us with an example which generalizes in an obvious way to
the Galois fields $GF(p^n)$ (where $p$ is a prime and $n \in \Nl$).
Though in this case we leave the usual umbral framework since we
consider a field which is not of characteristic zero, some basic
constructions and results remain valid.        \hfill {\Large $\Box$}
\vskip.3cm

As long as we restrict our considerations to operators acting on
polynomials, everything works smoothly. We are, however, also
interested in more general classes of functions and in
particular power series. In general, an umbral operator
like $\,\tilde{} \,$ does not preserve convergence of such a series.
The result of the application of an umbral operator to a power
series a priori only makes sense as a {\em formal} power series.
A problem is then to determine its domain of convergence (which
may be empty) and a possible continuation.
It seems that little is known about the convergence of power series
obtained via umbral maps.

\section{Symmetry operators on a lattice: an example}
\label{Solatt}
\setcounter{equation}{0}
In this section we generalize the example 3 of section \ref{IntroUC}
to $n$ dimensions. As an application of the umbral method, a
representation of the Lie algebra of $SO(3)$ on a lattice is then
presented. Let $x_1, \ldots , x_n$ be the canonical coordinate
functions on an n-dimensional (hypercubic) lattice with spacings
$a_i$. We define delta operators
\begin{eqnarray}
 (Q_i f)(x) := {1 \over a_i} \, \lbrack f(x_1, \ldots, x_{i-1}, x_i
               +a_i, x_{i+1}, \ldots, x_n) - f(x) \rbrack
\end{eqnarray}
acting on functions of $x=(x_1, \ldots, x_n)$.
The corresponding Pincherle derivatives are the shift operators $S_i$
acting on functions as follows,
\begin{eqnarray}         \label{shift-ops}
 (S_i f)(x) := f(x_1, \ldots, x_{i-1}, x_i+a_i, x_{i+1}, \ldots, x_n)
               \; .
\end{eqnarray}
The operators $\hat{\bf x}_i = {\bf x}_i \, S_i^{-1}$ and $Q_j$
then satisfy the commutation relations (\ref{ccr}) on the
algebra of polynomials in the variables $x_1, \ldots, x_n$.
\vskip.2cm

As outlined in the introduction, given a representation of a Lie
algebra in terms of the operators ${\bf y}_i$ and
$\partial/\partial y_j$ acting on functions on $\Rl^n$, (\ref{umap})
maps it into a representation by operators acting on functions on a
lattice. For the angular momentum operators in three dimensions this
means
\begin{eqnarray}
    L_i = -{\rm i} \sum_{j,k} \epsilon_{ijk} \, {\bf y}_j \,
                {\partial \over \partial y_k}
    \quad  \mapsto \quad
    \tilde{L}_i := - {\rm i} \sum_{j,k} \epsilon_{ijk} \, \hat{\bf x}_j
                   \, Q_k
\end{eqnarray}
where
\begin{eqnarray}
     \tilde{L}_i f(x)
   = - {\rm i} \, \sum_{j,k} \epsilon_{ijk} \, x_j \, (Q_k f)(x-a_j)
\end{eqnarray}
using the notation $x-a_j = (x_1, \ldots, x_{j-1}, x_j - a_j, x_{j+1},
\ldots, x_n)$.
\vskip.2cm

What are the corresponding `spherically symmetric' functions on the
lattice ? We have to find the solutions of $\tilde{L}_i f(x) =0$. From
the corresponding solution in the continuum case, we know that $f$
should depend on $x_k$ only through $\sum_{k=1}^3 \, \hat{\bf x}_k^2 1
= \sum_{k=1}^3 \, x_k (x_k - a_k)$ or $\ast$-products of this
expression. The set of lattice points determined by the equation
$ \sum_{k=1}^3 \, x_k (x_k - a_k) = constant \, $
therefore constitutes the analogue of the $2$-sphere in the
continuum case. Of course, only for special values of the constant
it will be non-empty. For a lattice with equal spacings in all
dimensions, the mappings $x_k \leftrightarrow x_\ell$ and
$x_k \mapsto a - x_k$ leave the above expression invariant and thus
help to construct the `lattice spheres'.

\section{Isospectral discretization of eigenvalue equations via umbral
         calculus ?}
\label{isospectral}
\setcounter{equation}{0}
In \cite{Turb92} differential equations were called `quasi-exactly
solvable' if there is at least one polynomial solution and `exactly
solvable' if there is a complete set of polynomial solutions.
The relevance for physics has been established in a series of papers
\cite{T} where quantum mechanical eigenvalue problems were collected
which can be reduced to equations having polynomial eigenfunctions
via an {\em ansatz} of the form
\begin{equation}
             \psi(y) = \phi(y) f(y)
\end{equation}
with a fixed non-polynomial function $f$ on $\Rl$. The most familiar
example is provided by the (one-dimensional) harmonic oscillator.
In this case
\begin{equation}
          \psi(y) = \phi(y) \, e^{- y^2/2}
\end{equation}
converts the Schr\"odinger equation into a differential equation for
$\phi$ which has the Hermite polynomials as a complete set of solutions.
Another example is the radial part of the Schr\"odinger equation for
a hydrogen atom.
\vskip.2cm

In \cite{ST95} a discretization procedure has been proposed for a
differential operator eigenvalue equation possessing polynomial
solutions such that the resulting difference equation has the same
spectrum. It corresponds to an umbral map in the sense of section
\ref{IntroUC} with the choice $Q = \partial_+$, the forward
discrete derivative operator.\footnote{The umbral framework provides
us with several alternatives, of course, which have not been
considered in \cite{ST95}.}
However, the procedure does not work well, in general, when applied to
the original eigenvalue problem which we started with. Though we do get
a discrete eigenvalue problem in this way which is naively\footnote{In
order to formulate a well-defined eigenvalue problem, we have to specify
a suitable function space in which we are looking for solutions. Each
eigenfunction of a differential operator is mapped to an eigenfunction
of the corresponding discrete operator (or at least a formal power
series which satisfies the discrete eigenvalue equation). Note, however,
that the discrete equation may have additional solutions. In
particular, this is the origin of boson or fermion doubling
in lattice field theories (cf \cite{S85,MM94}). }
isospectral, problems arise when we try to translate the
non-polynomial solutions. This will be illustrated with the following
examples.
\vskip.3cm
\noindent
{\em Example 1.} We apply the umbral map to a simple differential
equation,
\begin{eqnarray}             \label{exp_umap}
 {d \over dy} f(y) = k \, f(y) \quad \mapsto \quad
  Q \tilde{f}(x) = k \, \tilde{f}(x)  \; .
\end{eqnarray}
{}From the solution $f(y) = \exp(k y)$ of the differential equation
the corresponding solution of the difference equation on the rhs of
(\ref{exp_umap}) is then obtained as follows,
\begin{eqnarray}           \label{tildef_k}
   \tilde{f}(x) = f(\hat{\bf x}) \, 1
 = \sum_{\ell=0}^\infty \, {k^\ell \over \ell !} \, x^{(\ell)}  \; .
\end{eqnarray}
Though we would like to choose
$x$ as the canonical coordinate function on the lattice $a \Ir$, it may
be helpful at this point to consider it as a coordinate function
on $\Rl$ in view of a possible analytic continuation of the power
series obtained from the umbral procedure. A priori, we obtain
$\tilde{f}$ only as a {\em formal} power series. For {\em real} $k$, the
series in (\ref{tildef_k}) (which is a special case of a {\em Newton
series}) converges everywhere on the real line if $|k a| < 1$. For
$|k a| > 1$ the series is everywhere divergent, except for non-negative
integer multiples of $a$ (see \cite{Miln65}, for example).
The difference equation on the rhs of (\ref{exp_umap}) has no
nonvanishing solution for $k = -1/a$. For all other values of $k \in
\Cx$ the solutions are given by
\begin{eqnarray}
           \tilde{f}(n a) = \tilde{f}(0) \, (1 + k \, a)^n   \; .
\end{eqnarray}
With $\tilde{f}(0)=1$ this extends the series obtained above (for
$k > -1/a$).                \hfill {\Large $\Box$}
\vskip.3cm
\noindent
{\em Example 2.} Let us now apply the umbral map to the Hamiltonian of
the harmonic oscillator,
\begin{eqnarray}
   H = - {1 \over 2} \, {d^2 \over dy^2} + {1 \over 2} \,
             {\bf y}^2 \quad \mapsto \quad
   \tilde{H} = - {1 \over 2} \, Q^2 + {1 \over 2} \, \hat{\bf x}^2
   \; .
\end{eqnarray}
The eigenvalue equation for $H$ is then translated into the following
eigenvalue equation for $\tilde{H}$,
\begin{eqnarray}         \label{ho_diff}
   \tilde{H} \, \tilde{\psi}(x) =
 {1 \over 2} \, \lbrack - Q^2 \tilde{\psi}(x) + x (x-a) \,
 \tilde{\psi}(x-2a) \rbrack = E \, \tilde{\psi}(x)
\end{eqnarray}
which is a difference equation ($Q= \partial_+$).
{}From the solution $\psi_0(y)=\exp(-y^2/2)$ of the original eigenvalue
problem we obtain the solution
\begin{eqnarray}                   \label{gs_series}
  \tilde{\psi}_0 (x) = \sum_{\ell=0}^\infty \, {(-1)^\ell \over 2^\ell
                         \, \ell !} \, x^{(2 \ell)}
\end{eqnarray}
of the difference equation (\ref{ho_diff}) with $E=1/2$ as a formal
power series. Using
\begin{equation}
 {x^{(2 \ell+2)} \over 2^{\ell+1} \, (\ell+1)!} =
 {(x - 2 \ell a) (x - 2 \ell a - a) \over 2(\ell+1)} \,
 {x^{(2 \ell)} \over 2^\ell \, \ell !}
\end{equation}
the quotient criterium shows that the series is everywhere divergent,
except at values of $x$ which are non-negative integer multiples of
$a$ (where the series terminates). (\ref{gs_series}) thus only
determines a solution of the difference equation
\begin{equation}                \label{sing_de}
   x \, (x-a) \, \tilde{\psi}(x-2a) = {1 \over a^2} \lbrack
   \tilde{\psi}(x) - 2 \, \tilde{\psi}(x+a) + \tilde{\psi}(x+2a)
   \rbrack + \tilde{\psi}(x)
\end{equation}
on the non-negative part of $a \Ir$. The lhs of (\ref{sing_de})
vanishes for $x=0$ and $x=a$. Also the rhs vanishes if we calculate the
corresponding values of $\tilde{\psi}$ using (\ref{gs_series}). Our
solution can therefore be extended to the whole of $a \Ir$. But the
extension is not unique since $\tilde{\psi}(-a)$ and $\tilde{\psi}(-2a)$
can be chosen arbitrarily. This shows that the difference equation
has more independent solutions than the differential equation we
started with. The umbral-mapping of $\psi_0$ can, however, be
completed to yield a solution of the difference equation which exists
everywhere on $a \Ir$. This is done by expanding $\psi_0$ into
power series about negative multiples of $a$ and acting with $\,
\tilde{ } \,$ on these series.

The higher eigenfunctions of the harmonic oscillator are products of
Hermite polynomials with $\psi_0$,
\begin{eqnarray}
 \psi_n(y)= {\cal H}_n(y) \, \psi_0(y) \; .
\end{eqnarray}
Now $\tilde{\psi}_n(x)$ is obtained by replacing the ordinary product
by the $\ast$-product (cf the appendix), $\psi_0(y)$ by
$\tilde{\psi}_0 (x)$ as given above, and the Hermite polynomials by
the `discrete Hermite polynomials'. The latter are obtained from the
generating function
\begin{eqnarray}
 \tilde F(x,s) &=& \sum_{\ell=0}^{\infty} \, {\tilde {\cal H}_{\ell}(x)
 \over \ell !} \, s^{(\ell)} = \sum_{\ell=0}^{\infty} \sum_{k=1}^{\ell}
 {2^{\ell -k}(-1)^k \over k! \, (\ell -k)!} \, s^{(\ell +k)} \,
 x^{(\ell -k)}
\end{eqnarray}
by $\tilde{\cal H}_n(x) = (d/ds)^n \tilde{F}(x,s)|_{s=0}$.
                              \hfill {\Large $\Box$}

\section{Some more umbral calculus}
\label{modUC}
\setcounter{equation}{0}
There is a generalization of the calculus described in
section \ref{IntroUC}. Given a representation of the commutation
relation (\ref{ccr1}) by operators $Q$ and $\hat{\bf x}$ as in section
\ref{IntroUC}, and given an operator $A$ on the space of polynomials
which commutes with $Q$, then the new operator $\hat{\bf x} + A$
together with $Q$ also satisfies the commutation relation. In the
following, let $\hat{\bf x}$ denote such a more general choice (than
the special one in (\ref{umbral-hatx})). Defining
\begin{equation}
                  s_k(x) := \hat{\bf x}^k \, 1
\end{equation}
one finds
\begin{eqnarray}
                  Q s_k = k \, s_{k-1}
\end{eqnarray}
by use of the commutation relation (\ref{ccr1}). Such a polynomial
sequence $s_k$ is called a {\em Sheffer set} for the delta operator
$Q$ in the umbral literature. The basic polynomial
sequence $q_k$ for a delta operator $Q$ is a special Sheffer set.
If $s_k$ is a Sheffer set for $Q$, then there is an invertible
shift-invariant operator which maps the Sheffer polynomials $s_k$ to
the basic polynomials $q_k$. Furthermore,
\begin{eqnarray}
    s_n(x) = \sum_{k = 0}^n {n \choose k} s_k(0) \, q_{n-k}(x) \; .
\end{eqnarray}
We refer to \cite{RKO73} for proofs and further results. Again,
we define $\tilde{f}(x) := f(\hat{\bf x}) \, 1$.
\vskip.2cm

We have to stress that not all umbral results established for the
special choice (\ref{umbral-hatx}) for $\hat x$ translate to the more
general case considered in this section. In general, $s_k(0) \neq 0$
and the $s_k$ are not binomial.
\vskip.2cm

A particularly interesting choice for $\hat{\bf x}$ turns out to be
\begin{equation}           \label{hatx}
 \hat{\bf x} = \frac{1}{2} ( {\bf x} \, {Q'}^{-1} + {Q'}^{-1} \,
          {\bf x} )  \, .
\end{equation}
{}From umbral calculus we know that ${Q'}^{-1}$ commutes with $Q$. One
can then easily verify that $\lbrack Q , \hat{\bf x} \rbrack = \idty$.
The advantage of (\ref{hatx}) over (\ref{umbral-hatx}) is that it is
more symmetric and thus opens the chance to turn $\hat{\bf x}$ and
${\rm i} Q$ into Hermitian operators on a Hilbert
space.\footnote{If ${\bf x}$
is Hermitian and $Q$ anti-Hermitian, then $Q'$ and $Q'^{-1}$ are
Hermitian and thus also the operator in (\ref{hatx}).}
For $Q = d/dx$ we have $s_k(x) = q_k(x)$. In case of the
(forward) discrete derivative operator one finds
\begin{eqnarray}
 s_k(x) = \frac{1}{2^k}({\bf x} S_a^{-1}+S_a^{-1} {\bf x} )^k \, 1
        = (x-\frac{1}{2}\, a)(x-\frac{3}{2}\, a) \cdots
          (x - \frac{2k-1}{2} \, a )  \; .
\end{eqnarray}
Another realization of (\ref{hatx}), involving the central
difference operator, will be the subject of the following examples and
the next section. In that case, we have
\begin{eqnarray}
     s_0(x) = 1 \, , \quad s_1(x) = x \, , \quad s_2(x) = x^2 -{a^2
     \over 2} \, , \quad \ldots
\end{eqnarray}
using (\ref{central-Q'}) and $Q 1 = 0$.
\vskip.3cm
\noindent
{\em Example 1.}
Let us consider again the example of the harmonic oscillator. Using
(\ref{hatx}) and the central difference operator (\ref{centraldiff}),
the corresponding Schr\"odinger equation is umbral-mapped to
\begin{eqnarray}            \label{uSchroed}
 {\rm i} \, {\partial \over \partial t} \tilde{\psi}(x) =
   {1 \over 2} \, \lbrack - Q^2 +
   Q'^{-2} ({\bf x}^2 - {a^2 \over 2}) + 2 a^2 \, Q'^{-3} Q {\bf x}
   + {5 \over 4} a^4 \, Q'^{-4} Q^2 \rbrack \, \tilde{\psi}(x)
\end{eqnarray}
where on the rhs we have naively commuted all the non-local operators
$Q'^{-1}$ to the left. Acting with $Q'^4$ on this equation results in
a finite difference equation (with respect to the space coordinates).
However, if we discretize the time\footnote{This can be achieved
via an umbral map, of course.} in order to solve the initial value
problem for the above equation on a computer, calculation of the wave
function at the next time step requires $Q'^{-1}$.
But to explore an equation of the type above numerically, we
have to use an approximation with a {\em finite} lattice.
Choosing periodic boundary conditions (i.e., a periodic lattice), there
are convenient formulas for $Q'^{-1}$. On a periodic lattice with
$N=2m$ sites where $m$ is odd, the equation
\begin{equation}          \label{Q'invfinite}
        {Q'}^{-1} = \sum_{k=0}^{m-1} (-1)^k \, S_a^{2k+1}  \; .
\end{equation}
holds.\footnote{For even $m$, $Q'$ is not invertible.}
For odd $N$ one finds instead
\begin{equation}
  {Q'}^{-1} = \sum_{k=0}^{(N-1)/2} (-1)^{k+(N-1)/2} \, S_a^{2k}
  + \sum_{k=0}^{(N-1)/2 -1} (-1)^k \, S_a^{2k+1}  \; .
\end{equation}
In the following section, the quantum mechanical setting behind
(\ref{uSchroed}) is investigated more rigorously.
                                 \hfill {\Large $\Box$}
\vskip.3cm
\noindent
{\em Example 2.}
In section \ref{Solatt} we determined the `lattice spheres' with
respect to some umbral representation. Instead of (\ref{umbral-hatx})
here we choose
\begin{eqnarray}             \label{cd_xhat}
  \hat{\bf x}_i := {\bf x}_i \, ( S_i + S_i^{-1})^{-1}
                   + ( S_i + S_i^{-1})^{-1} \, {\bf x}_i
\end{eqnarray}
with the shift operators defined in (\ref{shift-ops}). This means that
we consider (\ref{hatx}) generalized to several dimensions with central
difference operators
\begin{eqnarray}                 \label{cd_nd}
      Q_i = {1 \over 2 a_i} \, ( S_i - S_i^{-1})  \; .
\end{eqnarray}
Using $Q''_i = a_i^2 Q_i$ and $Q_i 1 = 0$, we find the following
equations for `lattice spheres' in three dimensions,
\begin{eqnarray}
  \sum_{i=1}^3 (\hat{\bf x}_i)^2 \, 1 =
  \sum_{i=1}^3 \lbrack (x_i)^2 - {a_i^2 \over 2} \rbrack
  = \mbox{constant}   \; .
\end{eqnarray}
A spherically symmetric potential on the lattice is then a function
which only depends on $\ast$-products of $\sum_{i=1}^3
\lbrack (x_i)^2 - a_i^2/2 \rbrack $.      \hfill {\Large $\Box$}
\vskip.3cm
\noindent
{\em Example 3.}
A familiar representation of the Poincar{\'e} algebra is
\begin{eqnarray}
 P_\mu = - {\rm i} \, {\partial \over \partial y_\mu} \, , \quad
   M_i = \sum_{j,k} \epsilon_{ijk} \, {\bf y}_j \,  P_k \, , \quad
   L_i = y_0 \, P_i - \kappa \, {\bf y}_i \, P_0  \; .
                      \label{Poincare-rep}
\end{eqnarray}
These operators act on functions on $\Rl^4$ (with canonical coordinates
$y_\mu$). The commutation relations are then preserved when we perform
in the expressions (\ref{Poincare-rep}) the substitutions
\begin{eqnarray}            \label{sym-map}
  {\bf y}_i \quad \mapsto \quad \hat{\bf x}_i \; , \qquad
  {\partial \over \partial y_i} \quad \mapsto \quad Q_i \; ,
   \qquad i=1,2,3
\end{eqnarray}
with the operators defined in (\ref{cd_xhat}) and (\ref{cd_nd}).
In this way we obtain a representation of the Poincar{\'e} algebra
on a lattice with spacings $a_i$ (and continuous time as long as $y_0$
and $P_0$ are kept unchanged).\footnote{See also \cite{FS95}.
A representation on a four-dimensional space-time
lattice is obtained by extending the map (\ref{sym-map}) to $y_0$ and
$\partial/\partial y_0$.}
The quadratic Casimir operator of the Poincar{\'e} algebra in
this representation is
\begin{equation}                        \label{casimir}
      C = - \partial_t^2 + \sum_k Q_k^2   \; .
\end{equation}
where $\partial_t := \partial/\partial x_0$.
There is, however, a drawback of the representation presented above
and also those given in \cite{FS95}. As pointed out in
\cite{S85}, the Klein-Gordon equation built with the operator
(\ref{casimir}) suffers from a boson doubling problem analogous to the
more familiar fermion doubling problem in lattice field theory (see
\cite{MM94}, for example). This leaves us with a Poincar{\'e}-invariant
theory with 8 species of bosons. If the time dimension is also
discretized, one obtains 16 species. The relation between this
Klein-Gordon equation and the Dirac equation for `naive lattice
fermions' is the same as in the continuum,
\begin{equation}
 ({\rm i} \, \gamma^0 \partial_t + {\rm i} \, \gamma^k Q_k - m) \,
 ({\rm i} \, \gamma^0 \partial_t + {\rm i} \, \gamma^k Q_k + m)
 = - \partial_t^2 + \sum_k Q_k^2 - m^2  \; .
\end{equation}
The representation of the Poincar{\'e} algebra acting on continuum
spinor fields is mapped via (\ref{sym-map}) to a
representation on the lattice which leaves the lattice Dirac
equation invariant.   \\
 \hspace*{1cm}                           \hfill {\Large $\Box$}

\section{Via umbral calculus to quantum mechanics on a lattice}
\label{QM}
\setcounter{equation}{0}
In this section we investigate the umbral discretization method with
the central difference operator $Q = (S_a - S_{-a})/(2a)$ and the
symmetric operator (\ref{hatx}). It will be shown that they define
selfadjoint operators on the Hilbert space $\ell_2(a \Ir)$, the space
of square summable functions on the infinite lattice with spacings $a$.
We thus have a rigorous framework to explore the `umbral map'.
\vskip.2cm

By standard arguments $\bf x$ is selfadjoint with domain $\lbrace f \in
\ell_2(a \Ir) \mid {\bf x} f \in \ell_2(a \Ir) \rbrace$. The Fourier
transformation  $f \mapsto F$ where
\begin{eqnarray}                   \label{Fourier}
 f(x) = {1 \over \sqrt{2 \pi}} \, \int_{-\pi/a}^{\pi/a} F(k) \,
        e^{{\rm i} k x}  \, dk
\end{eqnarray}
is an isomorphism $\ell_2(a \Ir) \rightarrow L^2_{\pi/a}$. Here and in
the following $L^2_b$ stands for $L^2( \lbrack -b , b \rbrack )$, the
space of square-integrable functions on the interval $\lbrack -b , b
\rbrack$. It is more convenient for us to define the domain of $\bf x$
now as follows,
\begin{eqnarray}
  {\cal D}_{\bf x} = \lbrace f \in \ell_2(a \Ir) \mid
   F \mbox{ absol. continuous}, \, F(-{\pi \over a}) = F({\pi \over a}),
   \, {dF \over dk} \in L^2_{\pi/a} \rbrace    \; .
\end{eqnarray}
The action of $\bf x$ on $\ell_2(a \Ir)$ then corresponds to the
action of ${\rm i} \, d/dk$ on the domain in $L^2_{\pi/a}$
specified above.\footnote{The latter is a standard textbook example of
a selfadjoint operator. Via Fourier transformation it is mapped to a
selfadjoint operator on ${\cal D}_{\bf x} \subset \ell_2(a \Ir)$.}
Its spectrum is $\lbrace n \, a \mid n \in \Ir \rbrace$.
\vskip.2cm

Next we note that $-{\rm i} Q$ is a bounded selfadjoint operator on
$\ell_2(a \Ir)$. In $L^2_{\pi/a}$ it acts by multiplication
with $\sin(ak)/a$. Concerning the umbral map we can conclude the
following,
\begin{eqnarray*}
  \begin{array}{l@{\quad}lcl}
  \mbox{operator:} &
   - {\rm i} \, d/dy  &  \quad \mapsto \quad &
   - {\rm i} Q \\
  \mbox{spectrum:} & \Rl &  &
  \lbrace \lambda \in \Rl \mid |\lambda| \leq {1 \over a} \rbrace \\
  \mbox{eigenfunctions:} &
  f_\lambda(y) = \exp({\rm i} \lambda y) & \mapsto &
  \tilde{f}_\lambda(x) = \exp\lbrack {\rm i} (x/a) \mbox{arcsin}
  (\lambda a) \rbrack
  \end{array}
\end{eqnarray*}
The eigenfunctions of $-{\rm i} Q$ can indeed be calculated directly
from the power series expansions for those of $-{\rm i} d/dy$ (with
the help of \cite{Rior}, section 6.5).
The spectrum of $-{\rm i} Q$ is bounded, however, in accordance with the
boundedness of the operator.
Only in the limit $a \to 0$ we recover the full
spectrum of the continuum momentum operator.
\vskip.2cm

For $f \in {\cal D}_{\bf x}$ we have $Qf \in {\cal D}_{\bf x}$. The
operator $Q' = \lbrack Q , {\bf x} \rbrack = (S_a + S_{-a})/2 \,$ is
then defined on ${\cal D}_{\bf x}$. It is bounded and can be extended
to a selfadjoint operator on $\ell_2(a \Ir)$. $Q' f = 0$ for
$f \in \ell_2(a \Ir)$ implies $f = 0$. Hence $Q'^{-1}$ exists
on ${\cal D}_{Q'^{-1}} = Q'(\ell_2(a \Ir))$ and is selfadjoint
(Lemma XII.1.6 in \cite{DS63}). The Fourier transform of $Q'$ acts in
$L^2_{\pi/a}$ by multiplication with $\cos(ak)$. The operator $Q'^{-1}$
therefore acts by multiplication with $1/\cos(ak)$ on the domain
$\lbrace F \in L^2_{\pi/a} \mid F(k)/\cos(ak) \in L^2_{\pi/a} \rbrace$.
\vskip.3cm

It remains to investigate the operator
$\hat{\bf x} = ({\bf x} Q'^{-1} + Q'^{-1} {\bf x})/2$ which
is Hermitian on the dense domain
\begin{eqnarray}                  \label{domain_x^}
      {\cal D}_{\hat{\bf x}}
   =  \lbrace f \in {\cal D}_{\bf x} \cap {\cal D}_{Q'^{-1}} \mid
      {\bf x} f \in {\cal D}_{Q'^{-1}}, \, Q'^{-1} f \in
      {\cal D}_{\bf x} \rbrace    \; .
\end{eqnarray}
Without any calculations we can immediately conclude the following. The
operator $\hat{\bf x}^2$ can be defined on a dense domain on which it
commutes with complex conjugation. According to Theorem XII.4.18
and Corollary XII.4.13(a) in \cite{DS63} this operator has
selfadjoint extensions. Let us recall a theorem due to Rellich and
Dixmier (see Theorem 4.6.1 in \cite{Putn67}).
\vskip.3cm
\noindent
{\em Theorem.} Let $\bf q$ and $\bf p$ be closed Hermitian operators
on a Hilbert space $\cal H$ such that  \\
(1) $\lbrack {\bf p} , {\bf q} \rbrack = - {\rm i}$ on a subset
$\Omega \subset {\cal D}_{\bf q} \cap {\cal D}_{\bf p}$ dense in
$\cal H$ which is invariant under $\bf q$ and $\bf p$, \\
(2) ${\bf p}^2 + {\bf q^2}$ on $\Omega$ is essentially selfadjoint. \\
Then $\bf p$ and $\bf q$ are selfadjoint and unitarily equivalent to a
direct sum of Schr\"odinger representations.
                              \hfill {\Large $\Box$}
\vskip.3cm
\noindent
An isomorphism $\ell_2(a \Ir) \cong L^2(\Rl)$ maps the operators
$\hat{\bf x}$ and $-{\rm i} Q$ to corresponding operators in $L^2(\Rl)$.
These operators cannot be unitarily equivalent to those of the
Schr\"odinger representation since the latter are both unbounded.
Besides (2), the operators $\hat{\bf x}$ (which has a closed Hermitian
extension \cite{DS63}) and $-{\rm i} Q$ fulfil all assumptions of the
last theorem. Taking into account that $(-{\rm i} Q)^2$ is selfadjoint
and bounded, it follows that $\hat{\bf x}^2$ is not essentially
selfadjoint. Together with our previous result this means that
$\hat{\bf x}^2$ has inequivalent selfadjoint extensions.
\vskip.2cm

We now turn to a closer inspection of the operator $\hat{\bf x}$
which, via Fourier transformation, is translated into the
operator
\begin{eqnarray}
   X :=  {{\rm i} \over 2} \, \left( {1 \over \cos(ak)} {d \over dk}
        + {d \over dk} {1 \over \cos(ak)}  \right)
\end{eqnarray}
with domain ${\cal D}_X \subset L^2_{\pi/a}$ determined
by (\ref{domain_x^}). This operator is singular at $k = \pm \pi/(2a)$
and functions in ${\cal D}_X$ vanish at these points. Assuming that
the latter also holds for functions in the domain of selfadjoint
extensions\footnote{An inspection of the adjoint of $X$ (which
contains all selfadjoint extensions) indicates that this holds
indeed for all selfadjoint extensions. For our purposes it is
sufficient to verify a posteriori that it holds for all the selfadjoint
extensions which we construct below.},
the eigenvalue problem for $X$ separates into two independent eigenvalue
problems, namely for the following two operators. \\
(a) $X^{(1)}$ is $X$ restricted to ${\cal D}_{X^{(1)}} := \lbrace F \in
L^2_{\pi/2a} \mid F(k)/\cos(ak) \mbox{ abs. cont.},
F(-\pi/(2a))=0=F(\pi/(2a)), XF \in L^2_{\pi/2a} \rbrace$ \\
(b) $X^{(2)}$ is $X$ restricted to ${\cal D}_{X^{(2)}} := \lbrace F \in
L^2_\cup \mid F(k)/\cos(ak) \mbox{ abs. cont.},
F(-\pi/(2a)) = 0 = F(\pi/(2a)), F(-\pi/a) = F(\pi/a), XF \in
L^2_\cup \rbrace$  where
$L^2_\cup := L^2(\lbrack - \pi/a,-\pi/(2a)\rbrack \cup \lbrack \pi/(2a),
\pi/a \rbrack)$.   \\
In both cases we perform a change of coordinate
\begin{eqnarray}       \label{p-k}
                p = {1 \over a} \, \sin(a k)  \; .
\end{eqnarray}
Then, with the separation
\begin{eqnarray}
                  F(k) = \sqrt{|\cos(ak)|} \; \chi(p)
\end{eqnarray}
we find for $\ell =1,2$,
\begin{eqnarray}
  (X^{(\ell)} F)(k) =  {\rm i} \, \sqrt{| \cos(ak) |} \; {d \over dp}
  \chi(p)   \; .
\end{eqnarray}
The two operators $X^{(\ell)}$ now both translate into the more familiar
one
\begin{eqnarray}            \label{d/dp}
  {\rm i} \, {d \over dp}  \quad \mbox{on} \quad
  \lbrace \chi \in L^2_{1/a} \mid
  \chi \mbox{ abs. cont.}, \; \chi(-{1 \over a}) = 0 =
  \chi({1 \over a}), \;
  {d \chi \over dp} \in L^2_{1/a} \rbrace  \; .
\end{eqnarray}
Let $F_{(1)}$ and $F_{(2)}$ denote the restrictions of $F \in
{\cal D}_X$ to $\lbrack -\pi/a , \pi/a \rbrack$
and $\lbrack -\pi/a , -\pi/(2a) \rbrack \cup \lbrack \pi/(2a) ,
\pi/a \rbrack$, respectively. Then
\begin{eqnarray}
    (F, F') = \int_{-\pi/a}^{\pi/a} F(k)^\ast \, F'(k) \, dk
    = (\chi_{(1)} , \chi'_{(1)}) + (\chi_{(2)} , \chi'_{(2)})
\end{eqnarray}
where
\begin{eqnarray}
  (\chi_{(\ell)} , \chi'_{(\ell)}) = \int_{-1/a}^{1/a}
  \chi_{(\ell)}(p)^\ast \, \chi'_{(\ell)}(p) \, dp
  \qquad (\ell = 1,2)  \; .
\end{eqnarray}
The selfadjoint extensions of the operator (\ref{d/dp}) are
known to be given by
\begin{eqnarray}
  D_\alpha = {\rm i} {d \over dp}  \quad \mbox{on} \quad
  {\cal D}_\alpha := \lbrace \chi \in L^2_{1/a} \mid
  \chi \mbox{ abs. cont.}, \, \chi({1 \over a}) =
  e^{2 \pi {\rm i} \alpha} \, \chi(-{1 \over a}), \,
  {d \chi \over dp} \in L^2_{1/a} \rbrace  \quad
\end{eqnarray}
where $\alpha \in \lbrack 0 , 1)$ \cite{RSI}. A complete orthonormal
set of eigenfunctions of $D_\alpha$ is
\begin{eqnarray}
  \chi^\alpha_n(p) := \sqrt{a \over 2} \; \exp\lbrack -{\rm i} \,
  (\alpha + n) \, \pi \, a \, p \rbrack \, ,  \qquad  n \in \Ir \, ,
\end{eqnarray}
and $D_\alpha$ has a pure point spectrum $\lbrace (\alpha + n) \,
\pi \, a \mid n \in \Ir \rbrace$. A selfadjoint extension of the
operator $\hat{\bf x}$ is now obtained by choosing any pair from the set
of operators $D_\alpha$. It then defines operators $X^{(1)}_{\alpha_1}$
and $X^{(2)}_{\alpha_2}$ and in this way a selfadjoint extension
$\hat{\bf x}_{\alpha_1,\alpha_2}$ of $\hat{\bf x}$ with spectrum
$\lbrace (\alpha_1 + n) \, \pi \, a \mid n \in \Ir \rbrace \cup \lbrace
(\alpha_2 + n) \, \pi \, a \mid n \in \Ir \rbrace$. The operator
$\hat{\bf x}_{\alpha_1,\alpha_2}$ has the complete set of
eigenfunctions
\begin{eqnarray}
   f^{(\alpha_1)}_{n,1}(x) = {1 \over 2} \sqrt{a \over \pi}
   \int_{-{\pi \over 2a}}^{\pi \over 2a} \sqrt{\cos(ak)} \, \exp\lbrack
   -{\rm i} \, (\alpha_1 + n) \, \pi \, \sin(ak) + {\rm i} \, k \, x
   \rbrack \, dk   \hspace{3.2cm} \\
   f^{(\alpha_2)}_{n,2}(x) = {1 \over 2} \sqrt{a \over \pi}
   \left( \int_{-{\pi \over a}}^{-{\pi \over 2a}}
   + \int_{\pi \over 2a}^{\pi \over a} \right)
   \sqrt{|\cos(ak)|} \, \exp\lbrack - {\rm i} \, (\alpha_2 + n) \, \pi
   \, \sin(ak) + {\rm i} \, k \, x \rbrack \, dk   \qquad
\end{eqnarray}
in $\ell_2(a \Ir)$.
For the umbral map we can draw the following
conclusions,
\begin{eqnarray*}
  \begin{array}{l@{\quad}lcl}
  \mbox{operator:} &
  {\bf y}  &  \quad \mapsto \quad & \hat{\bf x}_{\alpha_1,\alpha_2} \\
  \mbox{spectrum:} &
  \Rl &  &
   \lbrace (\alpha_\ell + n) \, \pi \, a \mid n \in \Ir, \; \ell =1,2
   \rbrace       \\
  \mbox{eigenfunctions:} &
  f_\lambda(y) = \delta(y-\lambda) &  & f_{n,1}^{(\alpha_1)},
  \, f_{n,2}^{(\alpha_2)}     \; .
  \end{array}
\end{eqnarray*}
Of course, in this case we have no method to calculate
eigenfunctions of $\hat{\bf x}_{\alpha_1,\alpha_2}$ directly from
the generalized eigenfunctions $\delta(y-\lambda)$ of the Schr\"odinger
operator $\bf y$.
\vskip.2cm

Slightly more complicated is the case of the operator $\hat{\bf x}^2$.
Following our treatment of the operator $\hat{\bf x}$ itself,
a set of two selfadjoint extensions of the operator $-d^2/dp^2$
determines a selfadjoint extension of $\hat{\bf x}^2$. The
domains of selfadjoint extensions of $-d^2/dp^2$ in $L^2_{1/a}$ have
the form
\begin{eqnarray}
  {\cal D}_{b.c.} = \lbrace \chi \in L^2_{1/a} \mid \chi \mbox{
  differentiable, } {d \chi \over dp} \mbox{ abs. cont.}, \,
  {d^2 \chi \over dp^2} \in L^2_{1/a}, \, b.c. \rbrace
\end{eqnarray}
where $b.c.$ stands for a certain choice of boundary conditions,
like $\chi(-1/a) = 0 = \chi(1/a)$ (see \cite{RSI} for other choices).
\vskip.3cm
\noindent
{\em Example.} Let us consider the equation ${\bf a} \psi_0 = \kappa
\, \psi_0$ where ${\bf a} = \partial/\partial y + {\bf y}$ is the
annihilation operator for the one-dimensional harmonic oscillator,
and $\kappa \in \Cx$. The umbral map replaces $\bf a$ by
$Q + \hat{\bf x}$. Choosing for $\hat{\bf x}$ a selfadjoint
extension, we have to consider
\begin{eqnarray}      \label{gs_eq}
  (Q + \hat{\bf x}_{\alpha_1,\alpha_2}) \, \tilde{\psi}_0
  = \kappa \, \tilde{\psi}_0  \; .
\end{eqnarray}
We write the Fourier transform of $\tilde{\psi}_0$ as $\Psi_0(k)
= \sqrt{|\cos(ak)|} \, \chi_0(p)$ with $p$ given by (\ref{p-k}),
separately on $\lbrack -\pi/(2a) , \pi/(2a) \rbrack$ and
$\lbrack -\pi/a , -\pi/(2a) \rbrack \cup \lbrack \pi/(2a) , \pi/a
\rbrack$. Now (\ref{gs_eq}) translates
on both subsets of $\lbrack -\pi/a , \pi/a \rbrack$ to
$(p + d/dp) \chi_0 = \kappa \, \chi_0$ with the solution
$\chi_0(p) = C \, \exp(\kappa p - p^2/2)$ where $C$ is a constant.
For $C \neq 0$ one finds $\chi_0 \in {\cal D}_\alpha$ with $\alpha =
- {\rm i} \, \kappa/(\pi a)$. This restricts $\kappa$ since $\alpha
\in \lbrack 0,1)$. Furthermore, $\alpha_1 = \alpha_2 = \alpha$.
Application of the (umbral-mapped) creation operator
$-Q + \hat{\bf x}_{\alpha,\alpha}$ to $\tilde{\psi}_0$ leaves
${\cal D}_{\hat{\bf x}_{\alpha,\alpha}}$ since $p \, \chi_0(p)$ is
not in ${\cal D}_\alpha$. The algebraic construction of the
eigenfunctions for the harmonic oscillator therefore does not
survive after the umbral mapping. The problem actually appears
already in rewriting the Hamiltonian as
\begin{eqnarray}
    \tilde{H}_\alpha
 := {1 \over 2} \, (-Q^2 + \hat{\bf x}_{\alpha,\alpha}^2)
  = {1 \over 2} \, (-Q + \hat{\bf x}_{\alpha,\alpha})(Q +
    \hat{\bf x}_{\alpha,\alpha}) + {1 \over 2} \, \lbrack Q ,
    \hat{\bf x}_{\alpha,\alpha} \rbrack   \; .
\end{eqnarray}
The point is that $\lbrack Q , \hat{\bf x}_{\alpha,\alpha} \rbrack =
\idty$ does not hold on the domain of $\hat{\bf x}_{\alpha,\alpha}$.
As a consequence, there is no simple relation between the spectra of
$\tilde{H}_\alpha$ and ${1 \over 2} \, (-Q + \hat{\bf x}_{\alpha,
\alpha})(Q + \hat{\bf x}_{\alpha,\alpha})$.

In the way described above, the eigenvalue problem for a selfadjoint
extension of the Hamiltonian $\tilde{H} = (-Q^2 + \hat{\bf x}^2)/2$
reduces in $p$-space to (twice) the eigenvalue problem for the
Hamiltonian of the ordinary harmonic oscillator restricted to the finite
interval $\lbrack -1/a , 1/a \rbrack$ with the respective boundary
conditions. A choice among the many different selfadjoint extensions of
$\tilde{H}$ should be determined by the specification of the physical
system (on the lattice) which we intend to describe. It is not obvious
for us, however, what a natural choice could be.
                                           \hfill  {\Large $\Box$}
\vskip.3cm

An interesting aspect of the representation of the canonical commutation
relations considered in this section is the fact that it is solely
composed of the two operators $\bf x$ and $Q$ which both receive a
physical meaning if we interprete $\ell_2(a \Ir)$ as the space of
functions on a (physical) space lattice. $\bf x$ is the position
operator and $-{\rm i} Q$ the natural candidate for the
momentum operator (see also \cite{DMH92}). This is
the basis for a discrete version of quantum mechanics.
Whereas ordinary quantum mechanics has a continuous position space,
discrete quantum mechanics lives on a lattice. Quantum mechanical
models on a lattice should then be modelled with the selfadjoint
operators $\bf x$ and $-{\rm i} Q$. These satisfy commutation relations
which are different from the canonical ones. Still missing is,
however, a general recipe to quantize a (discrete) mechanical system,
analogous to canonical quantization. But what is the meaning of the
representation given by $\hat{\bf x}$ and $-{\rm i} Q$ ? Basically it
just offers us a way to get, apparently, close to the results of
ordinary quantum mechanics within the framework of discrete
quantum mechanics. That this representation is not equivalent
to the Schr\"odinger representation means that, within the framework
of discrete quantum mechanics, we cannot reproduce ordinary
quantum mechanics rigorously, at least not in the way attempted
in this section. In fact, we have found rather drastic deviations,
in particular a kind of spectrum doubling, a familiar problem in
lattice field theories \cite{S85,MM94}.

\section{Conclusions}
\label{Conclusions}
\setcounter{equation}{0}
In this paper we have pointed out that there is an apparently widely
unknown mathematical scheme, called umbral calculus,
behind recent work \cite{ST95,FS95} on discretization of differential
equations and physical continuum models. Using several examples we have
discussed its prospects and shortcomings. By choosing delta operators
different from those used in these papers, alternative discretizations
can be obtained. They have not been worked out in detail yet.
\vskip.2cm

Discretization of a continuum theory breaks the continuous space-time
symmetries which play a crucial role in (non-gravitational) quantum
field theory. There have been attempts to find a discrete analogue
of space-time symmetries
for lattice theories such that essential features of the continuum
group structures are maintained. Discretizations of Lorentz
transformations were considered in \cite{Schi49}, for example. In
\cite{Macr81} the Poincar{\'e} group acts on an ensemble of lattices
(see also \cite{Yama89} for a related point of view). Umbral calculus
offers a different way to implement symmetries on lattices
(see also \cite{FS95}).
\vskip.2cm

Umbral calculus provides us with certain classes of representations
of the canonical commutation relations. It is therefore of
potential interest for quantum mechanics and quantum field theory.
Among the variety of umbral maps which we have at our disposal, the
one determined by (\ref{hatx}) with the central difference operator
is of special interest (see also \cite{FS95}).
In this case we have a representation of the
canonical commutation relations constructed from the position and
the momentum operator on a lattice. This suggested a kind of
embedding of ordinary quantum mechanics into a formalism for quantum
mechanics on a lattice and thus a discretization of quantum mechanical
systems which is different from conventional ones (see \cite{Zakh93},
for example). The representation of the canonical commutation relations
obtained in this way is, however, not unitarily equivalent to the
Schr\"odinger representation. As a consequence, the image of ordinary
quantum mechanics under the umbral map cannot reproduce the results of
the former rigorously. We revealed a kind of spectrum doubling
similar to what is known in lattice field theories. This may be
regarded as a negative feature. In any case, we believe that this
is an interesting example of a representation of the canonical
commutation relations by selfadjoint operators which is not equivalent
to the Schr\"odinger representation. Furthermore, our analysis sheds
some light on the work in \cite{FS95} where this representation
has been used. The umbral framework yields many more examples, of
course, which can be analyzed analogously to the example which we
selected in section \ref{QM}.

\vskip.3cm
\noindent
{\em Acknowledgment.} F. M.-H. is grateful to Klaus Baumann for some
helpful discussions.

\renewcommand{\theequation} {\Alph{section}.\arabic{equation}}

\begin{appendix}

\section{On the $\ast$-product associated with the forward
         discrete derivative delta operator}
\setcounter{equation}{0}
Let $x$ be the canonical coordinate function on a lattice with
spacings $a$. A function $f(x)$ for which the finite difference
analogue of the Taylor series expansion (Gregory-Newton formula)
exists can be written as follows,
\begin{eqnarray}               \label{f-F}
  f(x) = \sum_{n=0}^\infty \, f_n \, x^n
       = \sum_{k=0}^\infty \, F_k \, x^{(k)}
\end{eqnarray}
where we have used
\begin{eqnarray}
    x^n = \sum_{k=0}^n \, S(n,k) \, a^{n-k} \, x^{(k)}  \; .
\end{eqnarray}
The coefficients $S(n,k)$ are the Stirling numbers of the second kind
($S(n,0)=0$ when $n>0$, $S(n,n)=1$). The coefficients $F_k$ in
(\ref{f-F}) are given by
\begin{eqnarray}
  F_k = \sum_{m=0}^\infty \, f_{k+m} \, S(k+m,k) \, a^m  \; .
\end{eqnarray}
The equation (\ref{f-F}) can also be expressed as
\begin{eqnarray}
                   f(x) = \tilde{F}(x)
\end{eqnarray}
where
\begin{eqnarray}
  F(\hat{\bf x}) = \sum_{k=0}^\infty \, F_k \, \hat{\bf x}^k  \; .
\end{eqnarray}
The $\ast$-product of two functions of $x$ is then given by
\begin{eqnarray}
  (f \ast h)(x) = \widetilde{(FH)}(x) = \sum_{k,\ell=0}^\infty
   \, F_k H_\ell \, x^{(k+\ell)}
\end{eqnarray}
and the rhs can be written as a power series in $x$ with the help of
\begin{eqnarray}
   x^{(n)} = \sum_{k=0}^n \, s(n,k) \, a^{n-k} \, x^k
\end{eqnarray}
where $s(n,k)$ are the Stirling numbers of the first kind.

\end{appendix}


\small

\begin{thebibliography}{99}
\bibitem{Roma+Rota78} Roman S and Rota G-C 1978 The Umbral Calculus
 {\em Adv. Math.} {\bf 27} 95  \\
 Roman S 1984 {\em The Umbral Calculus} (San Diego, CA: Academic Press)
\bibitem{RKO73} Rota G-C, Kahaner D and Odlyzko A 1973
 Finite operator calculus {\em J. Math. Anal. Appl.} {\bf 42} \\
 Rota G-C 1975 {\em Finite Operator Calculus}
 (San Diego, CA: Academic Press)
\bibitem{Turb92} Turbiner A 1992 On polynomial solutions of differential
 equations {\it J.Math.Phys.} {\bf 33} 3989
 {\em J. Phys. A} {\bf 25} L1087; 1994 Quasi-exactly-solvable differential
 equations, to appear in CRC Handbook of Lie Group Analysis of
 Differential equations, Vol. 3, ed N Ibragimov (CRC Press)
\bibitem{ST95} Smirnov Y and Turbiner A 1995 Lie-algebraic discretization
 of differential equations {\em Preprint} IFUNAM FT 95-68
\bibitem{FS95} Frappat L and Sciarrino A 1995 Lattice space-time from
 Poincar\'e and $\kappa$-Poincar{\'e} algebras {\em Phys. Lett. B}
 {\bf 347} 28
\bibitem{LNR92} Lukierski J, Nowicki A and Ruegg H 1992 New quantum
 Poincar{\'e} algebra and $\kappa$-deformed field theory {\em
 Phys. Lett. B} {\bf 293} 344
\bibitem{Rior} Riordan J 1968 {\em Combinatorial Identities}
 (New York: Wiley)
\bibitem{T} Turbiner A and Ushveridze A 1987 Spectral singularities
 and quasi-exactly solvable quantal problems {\it Phys.Lett. A}
 {\bf 126} 181 \\
 Turbiner A 1988 Quasi exactly solvable problems and $sl(2,\Rl)$
 algebra {\it Comm.Math.Phys.} {\bf 118} 467;
 1992 Lie algebras and polynomials in one variable
 {\em J. Phys. A} {\bf 25} L1087
\bibitem{S85} Stacey R 1985 Spectrum doubling and double-valuedness
 {\it J.Math.Phys.} {\bf 26} 3172
\bibitem{MM94} Montvay I and M\"unster G 1994 {\it Quantum fields
 on a lattice} (Cambridge, Cambridge U Press)
\bibitem{Miln65} Milne-Thomson L M 1965 {\em The Calculus of
 Finite Differences} (London: MacMillan \& Co Ltd.)
\bibitem{DS63} Dunford N and Schwartz J T 1963 Linear Operators, Part
 II: Spectral Theory (New York: Wiley)
\bibitem{Putn67} Putnam C R 1967 Commutation properties of Hilbert
 Space Operators (Berlin: Springer)
\bibitem{RSI} Reed M and Simon B 1972 Methods of Modern Mathematical
 Physics I (Academic Press)
\bibitem{DMH92} Dimakis A and M\"uller-Hoissen F 1992 Quantum
 mechanics on a lattice and $q$-deformations {\em Phys. Lett.}
 {\bf B295} 242
\bibitem{Schi49}
 Schild A 1949 Discrete space-time and integral Lorentz
 transformations {\em Canadian J. Math.} {\bf 1} 29 \\
 Hill E L 1955 Relativistic theory of discrete
 momentum space and discrete space-time {\it Phys. Rev.} {\bf 100}
 1780 \\
 Arshansky R I 1982 Lorentz transformations on the
 lattice {\it Int. J. Theor. Phys.} {\bf 21} 121
\bibitem{Macr81} Macrea K I 1981 Rotationally invariant field theory
 on lattices. I. General concepts {\it Phys. Rev. D} {\bf 23} 886
\bibitem{Yama89} Yamamoto H 1989 Discrete spacetime and Lorentz
 invariance {\it Nucl. Phys. B (Proc. Suppl.)} {\bf 6} 154
\bibitem{Zakh93} Zakhar'ev B N 1993 Discrete and continuous quantum
 mechanics. Exactly solvable models {\em Sov. J. Part. Nucl.}
 {\bf 23} 603
\end{thebibliography}
\end{document}